\documentclass[aip,reprint,amsmath,amssymb]{revtex4-2}

\usepackage{graphicx} 
\usepackage{dcolumn}  
\usepackage{bm}       

\usepackage[utf8]{inputenc}
\usepackage[T1]{fontenc}
\usepackage{mathptmx}
\usepackage{etoolbox}
\usepackage{multirow}

\usepackage{xcolor}
\definecolor{mydarkgreen}{rgb}{0,0.4,0}

\usepackage[draft]{changes}
\setdeletedmarkup{\textcolor{red}{\sout{#1}}}

\usepackage[colorlinks=true,linkcolor=blue,urlcolor=cyan,citecolor=red,anchorcolor=blue]{hyperref}

\usepackage{siunitx}

\makeatletter
\def\@email#1#2{%
 \endgroup
 \patchcmd{\titleblock@produce}
  {\frontmatter@RRAPformat}
  {\frontmatter@RRAPformat{\produce@RRAP{*#1\href{mailto:#2}{#2}}}\frontmatter@RRAPformat}
  {}{}
}%
\makeatother

\definecolor{watermelon}{rgb}{0.89, 0.45, 0.51}
\definecolor{ocreRouge}{RGB}{178,74,32} 
\definecolor{ocreOrange}{RGB}{204, 119, 34}

\begin{document}

\title{Fast momentum-selective transport of Bose-Einstein condensates via controlled non-adiabatic dynamics in optical lattices}

\author{Raja Chamakhi}
\affiliation{LSAMA, Department of Physics, Faculty of Science of Tunis, University of Tunis El Manar, 2092 Tunis, Tunisia}

\author{Dana Codruta Marinica}
\affiliation{Universit\'e Paris-Saclay, CNRS, Institut des Sciences Mol\'eculaires d'Orsay, 91405 Orsay, France}

\author{Naceur Gaaloul}
\affiliation{Leibniz Universit\"at Hannover, Institut f\"ur Quantenoptik, Welfengarten 1, 30167 Hannover, Germany}

\author{Eric Charron}
\affiliation{Universit\'e Paris-Saclay, CNRS, Institut des Sciences Mol\'eculaires d'Orsay, 91405 Orsay, France}

\author{Mourad Telmini}
\affiliation{LSAMA, Department of Physics, Faculty of Science of Tunis, University of Tunis El Manar, 2092 Tunis, Tunisia}

\begin{abstract}
We present a detailed numerical study of a protocol for momentum-selective transport of a Bose–Einstein condensate (BEC) in a one-dimensional optical lattice, achieving narrow momentum distributions through controlled non-adiabatic dynamics. The protocol consists of non-adiabatic loading into the lattice, coherent acceleration using a symmetric trapezoidal acceleration profile, and non-adiabatic release into free space. Using the time-dependent Gross–Pitaevskii equation, we simulate the full sequence and analyze the role of non-adiabatic excitations on the final momentum distribution. We identify the intra-site breathing dynamics as the dominant mechanism governing spectral purity under fast loading conditions. By tracking the condensate's spatial width during the evolution, we demonstrate a direct correlation with the final momentum spread. A variational model based on a Gaussian ansatz quantitatively reproduces the observed dynamics and provides physical insight into the breathing mechanism. Our results reveal the existence of “magic” times,  \emph{i.e}, specific loading or acceleration durations synchronized with the breathing oscillation period, where quasi-monochromatic momentum distributions can be achieved even with loading times as short as \SI{100}{\micro\second}. In the tight-binding regime, this approach offers speedup factors of 3 to 6 compared to adiabatic protocols while maintaining high transfer fidelities, providing a practical route to coherent transport for quantum sensors operating under stringent timing constraints.
\end{abstract}

\maketitle

\section{Introduction}
\label{sec:intro}

The controlled manipulation of ultracold atomic gases in optical lattices has become a central topic in quantum technologies, particularly for applications in quantum simulation, high-precision metrology, and inertial sensing \cite{Bloch2005, Morsch2006, Bloch2012, Bongs2019}. Among these systems, Bose–Einstein condensates (BECs) loaded into optical lattices offer a versatile and tunable platform for implementing large momentum transfer (LMT) operations, which are crucial for long-baseline atom interferometers and tests of fundamental physics \cite{Battesti2004, Clade2006, McDonald2013}.  To optimize the performance of these instruments, it is essential to preserve momentum selectivity, since the fringe contrast is highly sensitive to the momentum purity of the atomic source, which otherwise complicates accurate phase extraction \cite{Siemss2023}. While adiabatic protocols are known to minimize internal excitations and ensure such selectivity \cite{Wu2000, Geiger2010}, they typically require durations of hundreds of microseconds to milliseconds, which conflicts with the stringent timing constraints of compact interferometers operating under free-fall limits \cite{Muentinga2013} or in mobile platforms subject to vibrations and rotations \cite{Barrett2016}.

Various alternative strategies have been developed to circumvent this issue. On one hand, LMT beam splitters based on multiphoton Bragg diffraction \cite{Mueller2008}, Bloch oscillations \cite{Battesti2004, Clade2006, McDonald2014}, or hybrid combinations \cite{Sugarbaker2013, Savoie2018, Asenbaum2020} have achieved good performance, with momentum transfers up to several tens of $\hbar k_L$ (where $k_L$ is the lattice wave vector), sometimes at the cost of complex pulse sequences and parameter sensitivity. Recent advancements in BECs within optical lattices have deepened our understanding of quantum transport, nonlinear dynamics, and momentum-resolved imaging. For example, high-resolution studies of condensate transport dynamics have highlighted the importance of controlled time-of-flight imaging in probing lattice band populations and coherent tunneling phenomena \cite{Dupont2023, Dionis2025}. These breakthroughs provide a robust framework for examining condensate breathing, collective modes, and adiabatic acceleration protocols.

Recent efforts combining Floquet engineering with optimal control techniques \cite{Glaser2015, Reitter2017, Koch_2022, Ansel_2024} represent a fundamentally different approach that has enabled robust and fast preparation of target states with impressive momentum transfer of the order of $600\,\hbar k_L$ \cite{Rodzinka2024}. While powerful, these optimal approaches usually involve nontrivial control sequences, and sometimes lack clear interpretative models. Similarly, shortcut-to-adiabaticity (STA) methods \cite{Chen_2010, Torrontegui_2013, DGO_2019} and their enhanced variants \cite{Whitty_2020} (eSTA) using optimized temporal sequences have demonstrated efficient transfer of Bose-Einstein condensates into specific bands of optical lattices \cite{Zhou_2018} and robust fast atomic transport with improved stability against systematic errors and noise \cite{Whitty_2022}, but require protocols with carefully tuned optimized ramps. Counter-diabatic driving techniques have also been explored in various contexts, including recent proposals in driven optical lattices \cite{Giergiel2025}.

In parallel, several lattice-based strategies have been developed to realize compact, high-fidelity interferometers beyond conventional schemes. Shaken-lattice interferometry, where the position of the lattice is modulated in time, implements interferometric sequences directly within the lattice potential and enables acceleration sensing with a sensitivity that grows with the interrogation time~\cite{Weidner2017}. Excited-band Bloch oscillations at specific ``magic'' lattice depths also provide highly efficient momentum transfer while reducing lattice-induced phase noise~\cite{McAlpine2020}. Multidimensional Bloch-band architectures in optical lattices have demonstrated vectorial inertial sensing, including two-dimensional Bloch oscillations and a 2D atomic Michelson interferometer~\cite{LeDesma2025}. A complementary approach is offered by continuously trapped interferometers in Floquet-engineered lattices, where ``magic'' band structures suppress intensity-related phase noise and enable flexible, programmable sensor designs~\cite{Chai2025}.

Another approach to improving transport efficiency was demonstrated by Cladé et al. \cite{Clade2017}, who showed that preparing atoms in a Wannier-Bloch state, rather than in a standard Bloch state, can dramatically enhance the efficiency of coherent momentum transfer. This was achieved either through adiabatic acceleration ramps or by applying compensating phase shifts to the lattice. In contrast to such optimization strategies, our work investigates the possibility of realizing momentum-selective matter-wave transport in the non-adiabatic regime without requiring state preparation or phase compensation.

We analyze a complete transport sequence composed of a non-adiabatic loading stage into a one-dimensional optical lattice, coherent acceleration using a symmetric trapezoidal ramp, and subsequent non-adiabatic release into free space. The objective is to identify operational regimes where the condensate evolves coherently toward a spectrally narrow final momentum distribution, referred to as a quasi-monochromatic distribution, despite the rapid dynamics and absence of adiabatic protection. To this end, we numerically simulate the condensate dynamics using the time-dependent Gross–Pitaevskii equation (GPE) and analyze the momentum distribution at the end of the protocol. Surprisingly, we find that despite the rapid loading and release, highly monochromatic momentum distributions can be achieved when the loading/release or acceleration duration matches specific values, referred to as magic times, which synchronize with internal breathing oscillations. At these magic times, more than 98\% of the population is concentrated in the target momentum class, with sideband populations at $\pm 2\hbar k_L$ suppressed below 1\%. The analysis is supported by a variational model and a Fourier-based interpretation of the final state, providing physical insight into the interplay between internal breathing modes and spectral selectivity.

This work continues a line of investigation aimed at developing precision-controlled matter-wave transport protocols for quantum sensing applications. In particular, previous studies from the group of Ernst Maria Rasel at the University of Hanover demonstrated that species-selective lattice-launch techniques enable high-fidelity acceleration of dual-species Bose–Einstein condensates, with direct relevance to tests of the Weak Equivalence Principle and dual-species interferometry \cite{Chamakhi2015, Chamakhi2016}. While these implementations relied on near-adiabatic sequences, the present study explores the diabatic regime, offering new insight into how internal condensate dynamics affect momentum-space coherence and the possible formation of quasi-monochromatic momentum distributions. Our protocol thus provides a practical route toward coherent matter-wave sources suitable as building blocks for future interferometric sensors operating beyond the adiabatic regime.

The remainder of this paper is organized as follows. In Section~\ref{sec:theory}, we present the theoretical framework and the numerical methods used to simulate the full transport protocol. Section~\ref{sec:protocol} describes the loading, acceleration, and release sequence in detail. In Section~\ref{sec:results}, we analyze the resulting momentum distributions and identify optimal transport regimes. Section~\ref{sec:model} introduces the variational model and compares its predictions to the GPE results. Section~\ref{sec:monochromaticity} provides a physical analysis of the onset of monochromaticity based on the breathing dynamics. Finally, Section~\ref{sec:conclusion} summarizes our findings.

\section{Theoretical Framework and Numerical Methods}
\label{sec:theory}

We consider a dilute Bose–Einstein condensate of $^{87}$Rb atoms trapped in a one-dimensional potential $V(x,t)$ with strong transverse confinement, practically achieved using tightly focused optical dipole traps \cite{GRIMM2000}. In our simulations, the transverse trapping frequency $\omega_\perp = 2\pi \times 485.4$\,Hz is much larger than the longitudinal frequency $\omega_x = 2\pi \times 1.37$\,Hz, yielding an aspect ratio $\omega_\perp / \omega_x \simeq 350$. Under these conditions, the transverse degrees of freedom remain frozen in their ground state, and the dynamics can be accurately described within the mean-field approximation by the effective one-dimensional Gross–Pitaevskii equation \cite{Olshanii1998, Salasnich2002, Pethick2008}.
\begin{equation}
i\hbar\,\partial_t\varphi(x,t) = \Big[
\frac{\hat{p}^2}{2m}
+ V(x,t)
+ N g_\mathrm{1D} |\varphi(x,t)|^2
\Big] \varphi(x,t)\,,
\label{eq:GPE}
\end{equation}
where $m$ is the atomic mass. The effective one-dimensional interaction strength, $g_\mathrm{1D} = 2\hbar^2 a_s / (m a_\perp^2)$, arises from dimensional reduction under tight transverse confinement, where $a_\perp = [\hbar/(m \omega_\perp)]^{1/2}$ is the transverse harmonic oscillator length and $a_s$ is the $s$-wave scattering length. The GPE\;\eqref{eq:GPE} is numerically solved using a Fourier pseudo-spectral method for spatial discretization \cite{Kosloff1983}, in which the wave function is represented on a uniform grid and spatial derivatives are computed via fast Fourier transforms, ensuring high accuracy for smooth solutions. Time evolution is performed using a second-order split-operator scheme \cite{Feit1983}, where the kinetic and potential contributions to the Hamiltonian are applied sequentially in momentum and position space, respectively, with the time step symmetrically split to maintain second-order accuracy. These are standard techniques widely used in computational quantum dynamics. Simulations typically employ spatial grids of up to $N_\text{grid}=2^{18}$ points with a grid spacing $\Delta x = 3.5$\,nm. The time step $\Delta t = 100$\,ns is chosen to accurately resolve dynamics at sub-microsecond timescales. The initial wavefunction $\varphi(x,t=0)$ corresponds to the condensate in a cigar-shape, quasi-one-dimensional harmonic trap and is obtained via imaginary time propagation \cite{Chiofalo2000, Bao2004}.

The lattice depth $V_0$ is ramped up while the harmonic trap is simultaneously switched off using smooth envelope functions, as described in detail in Section \ref{sec:loading_descr}. The transverse confinement remains provided by the optical dipole trap throughout the sequence. Moreover, since the condensate is distributed over more than 800 lattice sites, the occupation number per site remains low (approximately 20 atoms at most), resulting in weak interactions at the single-site level. This largely decouples the longitudinal and transverse degrees of freedom, further justifying the one-dimensional treatment. During transport, a time-dependent lattice acceleration is applied, and finally the lattice is ramped down. These steps are described in detail in the following section.

The numerical results provide direct access to both the spatial density profiles and the momentum distributions, enabling a detailed analysis of coherence, excitations, and transport efficiency.

\section{Transport Protocol}
\label{sec:protocol}

The full transport protocol consists of three stages: (i) loading the condensate into the optical lattice, (ii) accelerating the lattice to impart momentum, and (iii) releasing the condensate. The loading and release stages are governed by smooth, time-dependent modulation of the external potential.

\subsection{Loading into the Optical Lattice}
\label{sec:loading_descr}

Initially, the condensate is confined in a harmonic trap. The optical lattice is ramped on while the harmonic potential is simultaneously switched off. Both operations are described by complementary ramp functions
\begin{align}
f_{\uparrow}(t; t_S, t_L)   & = \sin^2 \left[ \frac{\pi(t - t_S)}{2 t_L}\right] \\
f_{\downarrow}(t; t_S, t_L) & = 1 - f_{\uparrow}(t; t_S, t_L)
\end{align}
defined in the time interval $t_S \leqslant t \leqslant t_S + t_L$, where $t_S$ is the start time of the ramp and $t_L$ its duration.

The total potential during the loading phase, i.e. between $t=t_S=0$ and $t=t_L$, is given by
\begin{equation}
V(x,t) = V_h(x)\,f_{\downarrow}(t; 0, t_L) + V_\mathrm{OL}(x)\,f_{\uparrow}(t; 0, t_L)\,,
\end{equation}
where
\begin{equation}
V_h(x) = \frac{1}{2}\,m\,\omega_x^2\,x^2\,,
\end{equation}
and
\begin{equation}
V_\mathrm{OL}(x) = V_0\,\cos^2(k_L x)\,.
\end{equation}
In this last expression, $k_L = 2\pi / \lambda_L$ is the lattice wave vector, $\lambda_L$ being the lattice wavelength. The ramp duration $t_L$ determines the degree of adiabaticity: longer ramps preserve the condensate from excitations, while short ramps may induce non-adiabatic vibrational excitations within each lattice site, specifically the breathing mode, that can be harnessed for momentum shaping.

\subsection{Lattice Acceleration}

\begin{figure}[t!]
\centering
\includegraphics*[width=0.99\columnwidth]{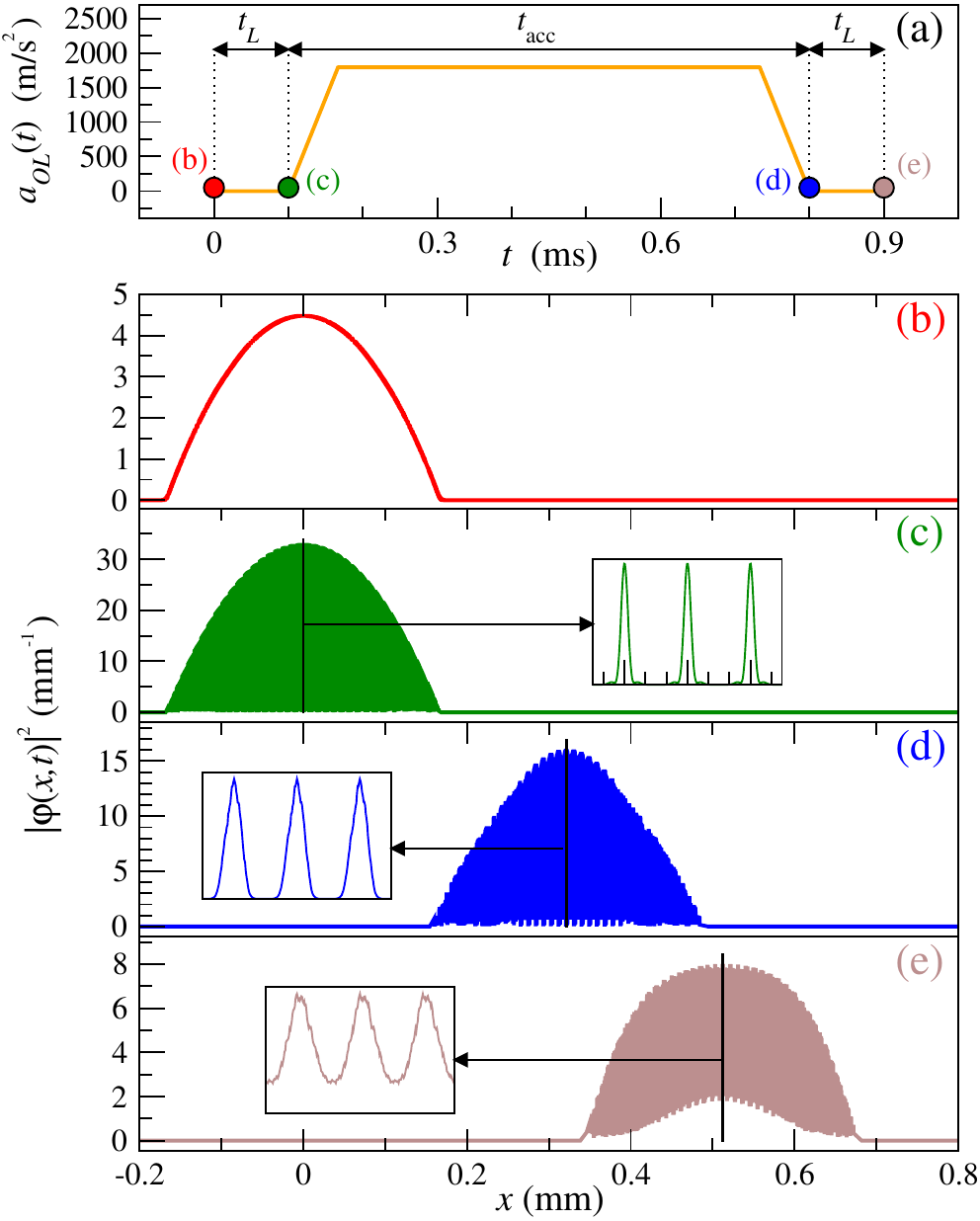}
\caption{Transport protocol and evolution of the condensate density.
(a) Time-dependent lattice acceleration $a_\mathrm{OL}(t)$ following a symmetric trapezoidal profile with total duration $t_{\mathrm{acc}} = 0.7$ ms and maximum value $a_{\mathrm{max}} = 1790$ m/s$^2$. Colored markers indicate the time points corresponding to subplots (b)–(e).
(b) Initial ground-state density of the BEC in the harmonic trap ($t = 0$), obtained for $N = 10^4$ atoms of $^{87}$Rb. The wave function is numerically  computed using the stationary Gross–Pitaevskii equation.
(c) Density profile after fast loading into the optical lattice ($t = t_L = 0.1$ ms). The lattice potential is ramped up while the harmonic trap is switched off over the same timescale. The inset shows the emergence of periodic density modulation at the lattice scale.
(d) Condensate density after the acceleration phase ($t = t_L + t_{\mathrm{acc}}$), illustrating the motion of the cloud 
induced by the lattice. The inset reveals persistent internal structure.
(e) Final density profile after a fast release from the lattice (also over a duration $t_L = 0.1$ ms). The central region remains relatively smooth, indicating small excitation.
All density profiles are plotted in position space. Insets in (c)–(e) zoom into the central part of the wave packet to highlight local density modulations induced by the lattice.}
\label{fig:fig1}
\end{figure}

Once loaded into the lattice at time $t=t_L$, the condensate is accelerated by translating the optical lattice according to a symmetric trapezoidal acceleration profile. This acceleration induces Bloch oscillations in quasi-momentum space: atoms in the fundamental band periodically traverse the first Brillouin zone, gaining $2 \hbar k_L$ of momentum each time they cross the zone boundary. The acceleration of the optical lattice starts at time $t_L$ and proceeds as follows: over a duration $\delta$, $a_\mathrm{OL}(t)$ increases linearly from zero to a maximum value $a_{\text{max}}$; it then remains constant at $a_{\text{max}}$ for an additional duration $\Delta$; finally, over another interval of duration $\delta$, the acceleration $a_\mathrm{OL}(t)$ decreases linearly back to zero. This symmetric temporal shape, shown in panel (a) of Fig.\,\ref{fig:fig1}, of total duration $t_\mathrm{acc}=2\delta+\Delta$, ensures controlled momentum transfer and we will see that it can also minimize unwanted sideband excitations at $\pm 2 \hbar k_L$ in the final momentum distribution. The constant-acceleration stage of duration $\Delta$, during which $a_\mathrm{OL}(t)=a_\mathrm{max}$, will be referred to as the plateau of the acceleration profile. The lattice speed $v_\mathrm{OL}(t)$ and position $x_\mathrm{OL}(t)$ can be easily obtained at any time $t$ by integrating the acceleration $a_\mathrm{OL}(t)$ once or twice, respectively. During the acceleration stage, the optical potential becomes
\begin{align}
V(x,t) = V_{0}\,\cos^2\!\Big(k_L\big[x-x_\mathrm{OL}(t)\big]\Big)\,.
\label{Vacc}
\end{align}
Note that the condensate speed may be different from the lattice speed in the non-adiabatic regime.

In the adiabatic limit, the number of Brillouin zone crossings (each corresponding to a momentum transfer of $2 \hbar k_L$) during the initial and final linear ramps is $N_\delta = a_{\text{max}} \delta / (2v_r)$, and during the plateau it is $N_\Delta = a_{\text{max}} \Delta / v_r$, where $v_r = \hbar k_L / m$ is the recoil velocity. In our simulations, we fix the total duration of the acceleration ramp $t_\mathrm{acc}=2\delta+\Delta$, and by carefully choosing $a_{\text{max}}$ and $\Delta$, we ensure that $N_\delta$ and $N_\Delta$ take integer values. This corresponds to an integer number of Bloch cycles (complete Brillouin-zone crossings), \emph{i.e.} the regime in which the Bloch sequence can map onto a single well-defined momentum class, which is the relevant situation for interferometric operation.

\subsection{Release from the Lattice}

After acceleration, the optical lattice is ramped down over a duration $t_R=t_L$ using the same envelope function $f_{\downarrow}$. The associated release potential reads
\begin{equation}
V(x,t) = V_0\,\cos^2\!\Big(k_L\big[x-x_\mathrm{OL}^f-v_\mathrm{OL}^ft\big]\Big)\,f_{\downarrow}(t; t_L+t_\mathrm{acc}, t_R)
\end{equation}
where $x_\mathrm{OL}^f$ and $v_\mathrm{OL}^f$ are the position and velocity of the lattice at the end of the acceleration ramp. Because of the symmetry ($t_R=t_L)$ of the trapezoidal acceleration ramp, as with the loading phase, the choice of $t_L$ determines the degree of adiabaticity during release.

\subsection{Unitary Transformation}

During the acceleration phase, it is useful to switch to the non-inertial reference frame that co-moves with the optical lattice. To achieve this, we apply a unitary transformation defined by the operator
\begin{equation}
\hat{U}_1(t) = e^{\frac{i}{\hbar}x_\mathrm{OL}(t)\,\hat{p}}\,,
\end{equation}
which corresponds to a spatial translation by $x_\mathrm{OL}(t)$, thereby shifting the system into the co-moving frame. The position and momentum operators transform as
\begin{subequations}
\begin{eqnarray}
\hat{U}_1 \, \hat{x} \; \hat{U}^{\dagger}_1 & = & \hat{x} + x_\mathrm{OL}(t)\,,\\
\hat{U}_1 \, \hat{p} \; \hat{U}^{\dagger}_1 & = & \hat{p}\,,
\end{eqnarray}
\end{subequations}
and the wavefunction in the new frame becomes
\begin{equation}
\Psi_1(x,t) = \hat{U}_1\,\varphi(x,t) = \varphi\big(x+x_\mathrm{OL}(t),t\big)\,.
\end{equation}
The corresponding nonlinear Hamiltonian, acting as the generator of the system’s dynamics, transforms as
\begin{eqnarray}
\hat{H}_1(t) & = & \hat{U}_1 \hat{H}\,\hat{U}^\dagger_1
+ i\hbar \frac{d\hat{U}_1}{dt} \hat{U}^{\dagger}_1\, \nonumber\\
& = & \frac{\hat{p}^2}{2m} + V_\mathrm{OL}(x)
+ Ng_\mathrm{1D}|\Psi_1(x,t)|^2 - v_\mathrm{OL}(t) \hat{p}\,.
\end{eqnarray}
A second unitary transformation is then applied to shift the momentum by $p_\mathrm{OL}(t)=mv_\mathrm{OL}(t)$
\begin{equation}
\hat{U}_2(t) = e ^{-\frac{i}{\hbar} p_\mathrm{OL}(t)\,\hat x}
\end{equation}
After this transformation, the position and momentum become
\begin{subequations}
\begin{eqnarray}
\hat{U}_2 \, \hat{x} \; \hat{U}^{\dagger}_2 & = & \hat{x}\,,\\
\hat{U}_2 \, \hat{p} \; \hat{U}^{\dagger}_2 & = & \hat{p} + p_\mathrm{OL}(t)\,.
\end{eqnarray}
\end{subequations}
The transformed wavefunction differs from $\Psi_1(x,t)$ by a position-dependent phase factor which does not affect the probability density
\begin{equation}
\Psi_2(x,t) = \hat{U}_2\,\Psi_1(x,t) = e^{-\frac{i}{\hbar} p_\mathrm{OL}(t)\,x}\;\Psi_1(x,t)\,.
\end{equation}
The resulting non-linear Hamiltonian is
\begin{equation}
\hat{H}_2(t) =
\frac{\hat{p}^2-p^2_\mathrm{OL}\!(t)}{2m}
+ V_\mathrm{OL}\!(x)
+ N g_\mathrm{1D} |\Psi_2(x,t)|^2
+ m a_\mathrm{OL}\!(t) \hat{x}
\end{equation}
Finally, a third unitary transformation
\begin{equation}
\Psi_3(x,t) = \hat{U}_3(t)\,\Psi_2 (x,t)
\end{equation}
can be applied to eliminate the purely time-dependent kinetic energy term of $\hat{H}_2(t)$ using
\begin{equation}
\hat U_3(t) = \exp\left(-\frac{i}{2m\hbar}\int_0^t p^2 _\mathrm{OL} (t^\prime)\,dt^\prime\right).
\end{equation}
This leads to the following form of the non-linear Hamiltonian
\begin{equation}
\hat H_3 (t) = \frac{\hat p ^2}{2m} + V_\mathrm{OL}(x) +Ng_\mathrm{1D} |\Psi_3(x,t)|^2 + m\,a_\mathrm{OL}(t)\,\hat{x}\,.
\label{eq:GPE2}
\end{equation}
It is worth noting that in the final Hamiltonian of Eq.\;(\ref{eq:GPE2}), the time dependence associated with the moving lattice potential has been transferred to a simpler time-dependent linear term $m\,a_\mathrm{OL}(t)\,\hat{x}$, representing the inertial force in the co-moving frame. The periodic potential $V_\mathrm{OL}(x)$ is now time-independent, which simplifies the numerical treatment. This approach allows us to compute the condensate evolution more efficiently while maintaining accurate tracking of internal excitations and momentum transfer during the acceleration sequence.

\section{Numerical Results}
\label{sec:results}

We simulate the full protocol described in Section~\ref{sec:protocol} using the time-dependent one-dimensional Gross–Pitaevskii equation\;(\ref{eq:GPE2}). The initial state is the ground state of $N = 10^4$ atoms of $^{87}$Rb in a harmonic trap with longitudinal frequency $\omega_x = 2\pi \times 1.37$\;Hz and transverse confinement $\omega_\perp = 2\pi \times 485.4$\;Hz. The scattering length is $a_s = 5.24$ nm. The optical lattice is formed by a laser with wavelength $\lambda_L = 768.96$\;nm, power $P = 4$\;W, and waist $w = 1$\;mm, resulting in a lattice depth $V_0 \simeq 104\;E_r$, where $E_r = \hbar^2 k_L^2 / (2m)$ is the recoil energy, and in our case $E_r / \hbar \simeq 24$\,kHz.

Figure~\ref{fig:fig1} summarizes the evolution of the condensate density throughout the protocol. Panel (a) shows the time-dependent acceleration $a_\mathrm{OL}(t)$ applied to the lattice over a total acceleration duration of $t_{\mathrm{acc}} = 0.7$ ms, following the trapezoidal profile described earlier. Panels (b) through (e) show the corresponding condensate density profiles, with the initial ground state in the harmonic trap in subplot (b), the density profile immediately after loading into the optical lattice at $t = t_L = 0.1$ ms in subplot (c), the density after acceleration to the final lattice velocity in subplot (d), and the final distribution after release from the lattice in subplot (e).

Each inset provides a zoomed view of the central region to illustrate the local density modulation induced by the lattice potential. These results show that despite the short duration of the loading and release ramps $(t_L = 0.1$\,ms), the condensate retains a well-defined, quasi-Gaussian density profile throughout the protocol.

\begin{figure}[t!]
\centering
\includegraphics*[width=0.99\columnwidth]{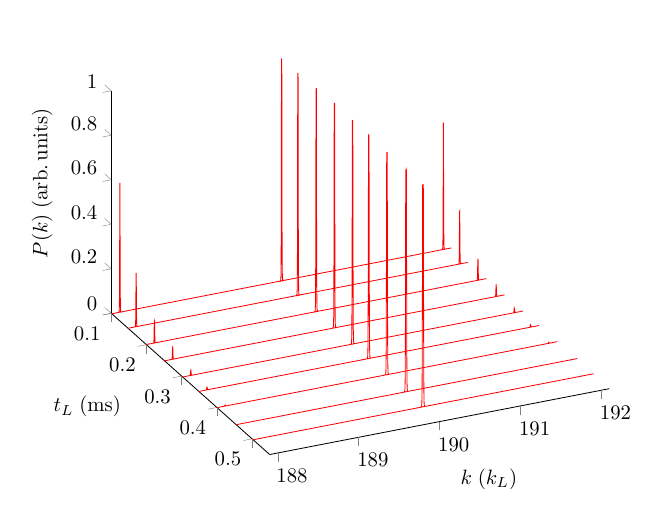}
\caption{Momentum distribution of the condensate after the transport protocol, as a function of the loading time $t_L$. Each curve represents the momentum distribution $P(k)$ (with the maximum of the central peak normalized to 1) computed for a fixed acceleration duration $t_{\mathrm{acc}} = 0.7$ ms and varying $t_L$ values from 0.1 ms to 0.5 ms. For short loading durations ($t_L \lesssim 0.3$ ms), non-adiabatic excitations are visible in the form of pronounced side peaks at $k = 188\,k_L$ and $k = 192\,k_L$ (i.e., $\pm 2 k_L$ from the main peak), reflecting population transfer into neighboring momentum states. As $t_L$ increases, these sidebands are progressively suppressed and the distribution becomes increasingly concentrated around $k = 190\,k_L$, the expected final momentum corresponding to the total number of momentum kicks set by the acceleration ramp. This transition illustrates the onset of adiabatic dynamics and the emergence of a narrow, spectrally pure momentum distribution.}
\label{fig:fig2}
\end{figure}

We now turn to the momentum-space representation of the final wave function. The momentum distribution $P(k)$ is computed as the squared modulus of the Fourier transform of the wave function $\varphi(x,t_f)$ at the end of the release stage, i.e. at $t_f=2t_L+t_\mathrm{acc}$, after the optical lattice has been fully ramped down. Specifically, we define
\begin{equation}
P(k) \propto \left| \int_{-\infty}^{+\infty} \varphi(x,t_f)\, e^{-i k x} \, dx \right|^2.
\end{equation}
This distribution captures the spectral content of the final condensate wave packet, which directly reflects the efficiency and coherence of the transport protocol. Figure~\ref{fig:fig2} displays the normalized momentum distribution $P(k)$ as a function of momentum $k$ (in units of $k_L$) and loading time $t_L$, with fixed acceleration duration $t_{\mathrm{acc}} = 0.7$ ms.

It is important to emphasize that the lattice acceleration parameters were chosen so that a total of $190$ momentum kicks are expected in the adiabatic limit. Therefore, observing a dominant peak centered around $k = 190\,k_L$ is consistent with this target but does not, by itself, indicate adiabaticity.

For short loading durations ($t_L < 0.3$ ms), significant population is transferred into neighboring momentum states ($\pm 2\hbar k_L$), located on either side of the target momentum class, indicating non-adiabatic excitations. The physical origin of this phenomenon, linked to intra-site breathing dynamics, is analyzed in detail in Sections \ref{sec:model} and \ref{sec:monochromaticity}. As $t_L$ increases, the side peaks are suppressed, and the distribution narrows around a central momentum $k = 190\,k_L$, reflecting adiabatic dynamics and the emergence of a narrow, spectrally pure momentum distribution.

\begin{figure}[t!]
\centering
\includegraphics*[width=0.99\columnwidth]{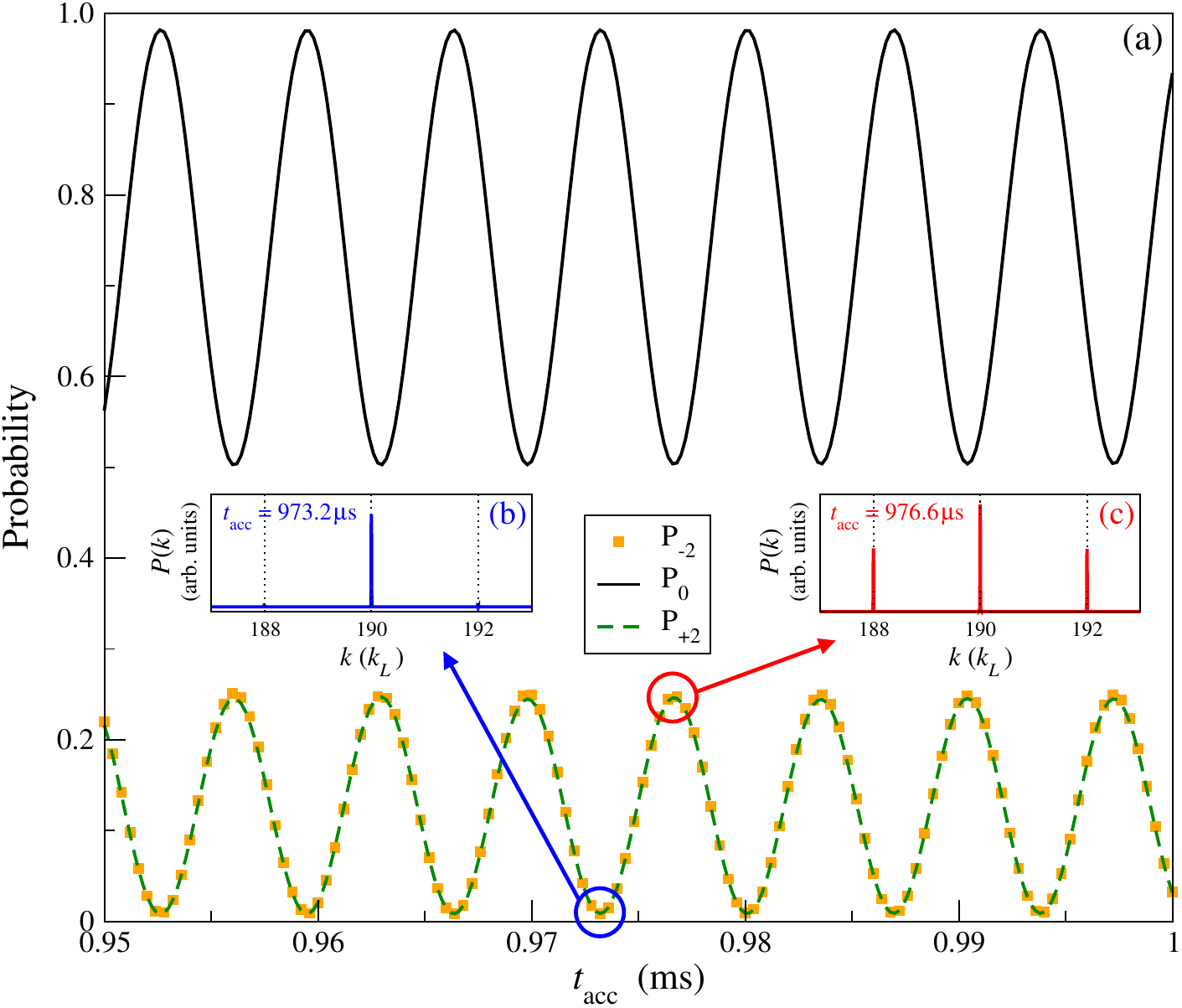}
\caption{(a) Momentum-state populations as a function of the acceleration time $t_{\mathrm{acc}}$, for a fixed loading duration $t_L = 0.1$ ms. The central peak population $P_0$ at $k = 190\,k_L$ (black curve) undergoes pronounced oscillations, periodically reaching values close to unity. The sideband populations $P_{-2}$ (orange squares) and $P_{+2}$ (green dashed line), corresponding to the first-order sideband peaks at $k = 188\,k_L$ and $192\,k_L$, oscillate out of phase with $P_0$, revealing coherent redistribution between momentum states as $t_{\mathrm{acc}}$ varies. Insets (b) and (c) show representative momentum spectra for two values of $t_{\mathrm{acc}}$: when $P_0$ is close to its maximum ($t_{\mathrm{acc}} = 973.2\ \mu$s, left inset b), the distribution is narrow and nearly monochromatic. In contrast, when $P_0$ is smaller ($t_{\mathrm{acc}} = 976.6\ \mu$s, right inset c), the sideband peaks are growing, illustrating the role of coherent non-adiabatic dynamics in shaping the final state.}
\label{fig:fig3}
\end{figure}

The transition toward adiabatic transport becomes clearly established for loading durations exceeding $0.4$\,ms, corresponding to approximately 60 times the characteristic breathing period of a lattice well ($\pi/\omega_{\mathrm{OL}} \simeq 6.3~\mu$s). In this regime, the side peaks in the momentum distribution are fully suppressed, and the population concentrates in a single, narrow momentum component. This confirms that sufficiently slow loading ensures minimal excitation and optimal spectral purity. The ability to prepare condensates in such well-defined momentum states is essential for high-fidelity quantum control in lattice-based experiments.

\begin{figure*}[t!]
\centering
\includegraphics*[width=1.75\columnwidth]{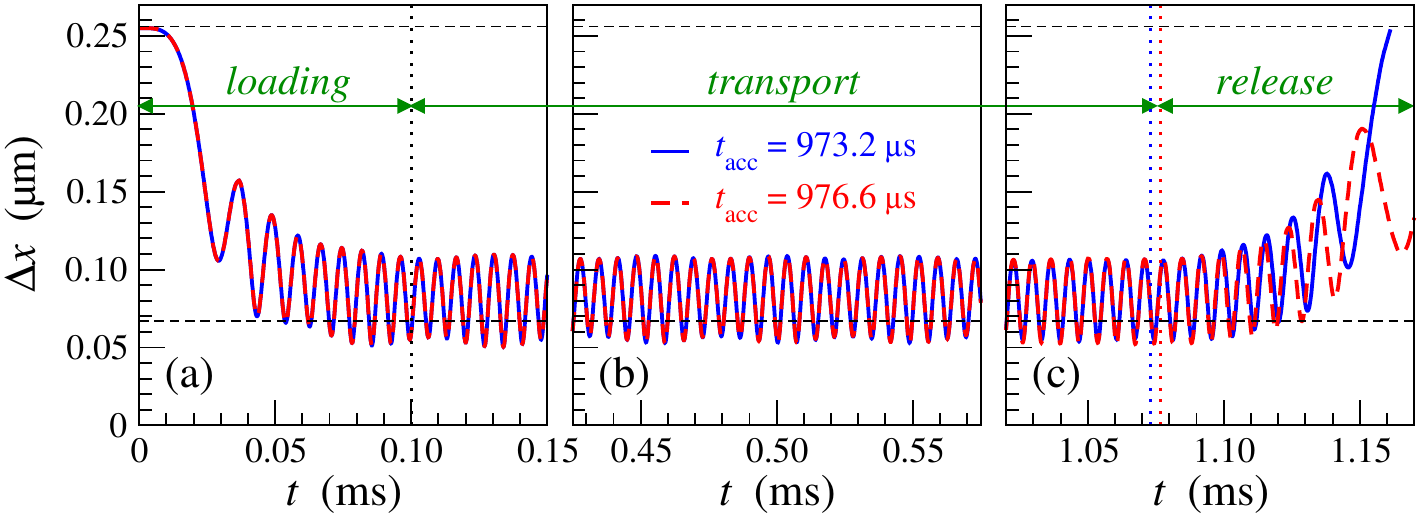}
\caption{Evolution of the intra-site spatial width $\Delta x(t)$ of the condensate wave function during the transport protocol, for fixed loading and release times $t_L = 0.1$\,ms and varying acceleration durations $t_{\mathrm{acc}}$. The width $\Delta x(t)$ is defined as the full width at half maximum (FWHM) of the density distribution, computed at each time step on a single central optical lattice site. Three successive stages are visible: (i) in panel (a) during loading $(t \leqslant t_L = 0.1\,\mathrm{ms})$, the condensate is compressed into the lattice wells, leading to a reduction in spatial extent; (ii) in panel (b) during the acceleration phase $(t_L \leqslant t \leqslant t_L+t_\mathrm{acc})$, $\Delta x(t)$ exhibits coherent oscillations, reflecting breathing-like motion within the lattice; (iii) in panel (c) during the release $(t > t_L+t_\mathrm{acc})$, the spatial width expands progressively. The two curves correspond to acceleration times $t_{\mathrm{acc}} = 973.2~\mu\mathrm{s}$ (solid blue line) and $t_{\mathrm{acc}} = 976.6~\mu\mathrm{s}$ (red dashed line). For both cases, $\Delta x(t)$ follows nearly identical dynamics during the loading and acceleration phases, indicating an identical coherent evolution. The amplitude of oscillations during the acceleration phase serves as a diagnostic of non-adiabatic excitations. A clear difference appears during the release phase, where the curves exhibit a phase shift and different final widths, reflecting dephasing and distinct expansion dynamics due to accumulated dynamical differences. The final value of $\Delta x(t)$ correlates with the spectral purity of the momentum distribution (see text for details).}
\label{fig:fig4}
\end{figure*}

To further investigate the dynamics in the non-adiabatic regime, we now analyze the momentum-state populations as a function of the acceleration time $t_{\mathrm{acc}}$, with the loading time fixed at $t_L = 0.1$ ms. This choice places the system well within the non-adiabatic regime, where coherent population transfer between neighboring moment states appear. The populations $P_0$, $P_{-2}$, and $P_{+2}$ are extracted from the final momentum distribution $P(k)$ by integrating around the corresponding peaks centered at $k = 190\,k_L$, $188\,k_L$, and $192\,k_L$, respectively. The integration windows are chosen to match the Brillouin zone width, ensuring clean separation between the modes. Higher-order momentum states were also investigated but their total population remains negligible (below $10^{-3}$), as shown in the inset of Fig.\;\ref{fig:fig7}, justifying our focus on the three principal momentum classes.

Figure~\ref{fig:fig3} shows the evolution of these populations as $t_{\mathrm{acc}}$ varies. The central component $P_0$ (solid black line) exhibits strong oscillations, periodically reaching values close to unity. This reflects the coherent buildup of population in the target momentum state, corresponding to the 190 momentum kicks imparted during acceleration. These oscillations have a characteristic period of approximately $6.85\,\mu$s, which is close to the harmonic breathing period in a single lattice well ($\pi/\omega_{\mathrm{OL}} \simeq 6.3\,\mu$s in the harmonic approximation for $V_0 \simeq 104\,E_r$). The small discrepancy between these two values arises from the anharmonicity of the actual cosine-squared lattice potential, which deviates from a purely harmonic well. This observation suggests that the dynamics are governed by intra-well breathing oscillations, consistent with the tight-binding regime realized in our system, with $V_0 \gg E_r$. Finally, the sideband populations $P_{\pm 2}$ (orange squares and green dashed line) oscillate out of phase with $P_0$, indicating coherent and tunable redistribution among neighboring momentum classes, governed by the ramp duration. From a practical standpoint, while the exact magic times depend on the specific protocol parameters and cannot be predicted analytically, the characteristic breathing period $\pi/\omega_{\mathrm{OL}}$ provides a predictable timescale that defines the range over which one needs to scan to identify the optimal durations experimentally.

Insets in Figure~\ref{fig:fig3} show two representative momentum distributions for values of $t_{\mathrm{acc}}$ where $P_0$ is either close to 1  (inset (b)) or strongly reduced (inset (c)). When $P_0$ is maximum, the distribution is spectrally narrow, centered on the desired final momentum. In contrast, when $P_0$ is minimum, significant population is transferred to the sidebands, indicating coherent non-adiabatic redistribution.

This oscillatory behavior could be tentatively interpreted as arising from quantum interferences between different components of the condensate wave function during the symmetric trapezoidal transport, an interpretation that will later be refined as the manifestation of a collective “breathing”-like dynamics within the lattice sites. As anticipated in the introduction, this result confirms that the system can exhibit quasi-monochromatic output even in the fast-loading and fast-release regime, provided that the acceleration duration $t_{\mathrm{acc}}$ is carefully tuned. The oscillatory structure visible in Fig.~\ref{fig:fig3} thus reveals the existence of several periodic magic times, \emph{i.e.} specific values of $t_{\mathrm{acc}}$ where $P_0$ reaches near-unity values despite the non-adiabatic loading. At these specific durations, destructive interference between the transiently populated $\pm 2 \hbar k_L$ momentum states effectively rephases the wave function into a narrow momentum distribution. This mechanism offers a practical route to spectral purity without adiabatic protocols.

To verify that the oscillations observed in Figure~\ref{fig:fig3} do not originate from a center-of-mass motion of the condensate, we computed the expectation value of the position $\langle x(t) \rangle$ in a single site throughout the full sequence. The result confirms that the condensate remains perfectly centered at the lattice site and faithfully follows the imposed lattice displacement, without exhibiting any global motion in the moving frame. This rules out center-of-mass oscillations as the origin of the oscillatory behavior in $P_0$. Instead, the observed dynamics must arise from coherent size oscillations, i.e., collective breathing modes within the lattice wells. This justifies the relevance of the intra-site spatial width $\Delta x(t)$ as a complementary observable to probe coherent excitation and spectral purity.

Figure~\ref{fig:fig4} provides insight into the internal dynamics of the condensate by tracking the intra-site spatial width $\Delta x(t)$ during the full transport protocol, for the two closely spaced acceleration durations $t_{\mathrm{acc}} = 973.2~\mu\mathrm{s}$ and $976.6~\mu\mathrm{s}$. These two cases correspond respectively to a maximum and a minimum of the central momentum population $P_0$, as previously discussed in Fig.~\ref{fig:fig3}.

During the loading stage and throughout most of the acceleration phase, the two trajectories remain virtually indistinguishable, exhibiting coherent breathing-like oscillations with the same amplitude and phase. This reflects the fact that the condensate undergoes nearly identical internal evolution up to the final portion of the acceleration stage, as the two values of $t_{\mathrm{acc}}$ differ only slightly. Consequently, the acceleration ramps closely overlap until the very end of the protocol. A clear difference emerges during the release stage, where the trajectories begin to dephase, resulting in distinct final values of $\Delta x(t)$. This divergence arises from subtle differences in internal excitations and accumulated phase, which affect the expansion dynamics once the lattice is switched off. Importantly, the trajectory corresponding to $t_{\mathrm{acc}} = 973.2~\mu\mathrm{s}$ leads to a broad final spatial profile, coinciding with a spectrally pure, nearly monochromatic momentum distribution. Conversely, the narrower final width observed for $t_{\mathrm{acc}} = 976.6~\mu\mathrm{s}$ reflects the presence of multiple momentum components.

In the next section, we introduce a simplified model that further supports the interpretation of $\Delta x(t)$ as a sensitive real-space indicator of spectral purity under non-adiabatic conditions.

\begin{figure*}[t!]
	\centering
	\includegraphics*[width=1.75\columnwidth]{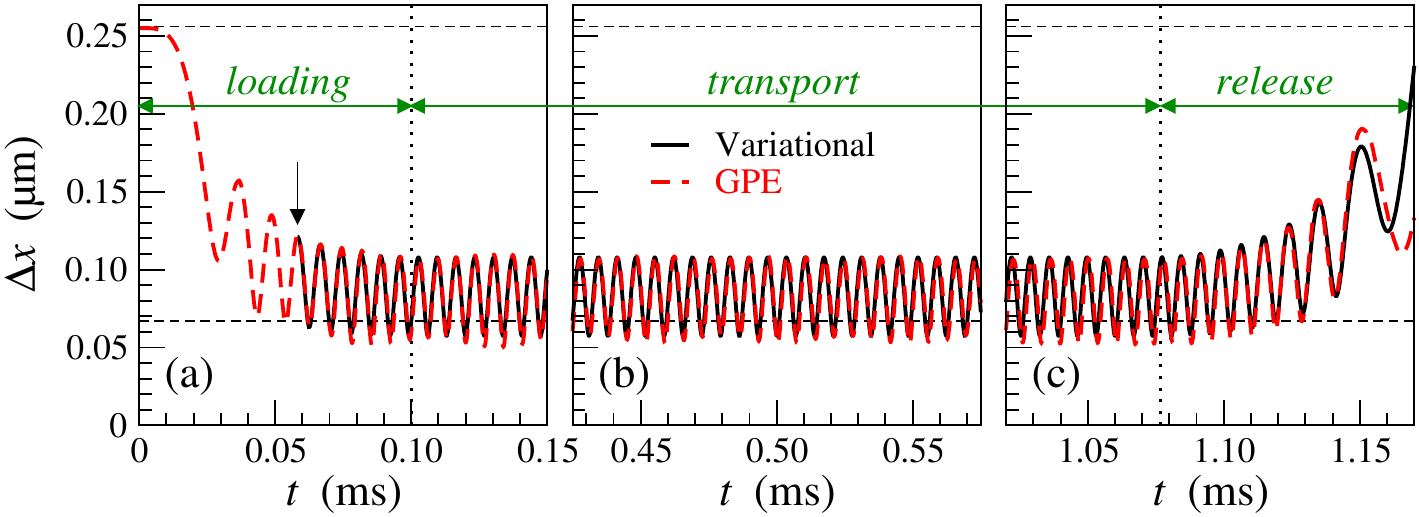}
	\caption{Comparison between the full numerical simulation and the variational model for the intra-site spatial dynamics of the condensate. The red dashed line shows the time-dependent width of the central density peak, $\Delta x(t)$, expressed as the full width at half maximum (FWHM), extracted from the Gross–Pitaevskii equation for $t_{\mathrm{acc}} = 976.6~\mu\mathrm{s}$. The black solid line corresponds to the evolution predicted by the variational model based on a Gaussian ansatz and assuming harmonic confinement within each lattice site. The variational propagation is initialized in panel (a) at the time $t_i$ of the fourth local maximum of $\Delta x(t)$, as obtained from the GPE solution. The corresponding value of $\sigma(t_i)$ is set from the numerical result to ensure accurate matching at the starting point. Excellent agreement is observed between the two approaches during the second half of the loading ramp in panel (a) and throughout the acceleration phase shown in panel (b), where the model captures both the amplitude and phase of the intra-site breathing oscillations. Discrepancies emerge at the end of the release stage in panel (c), when the wave function freely expands and begins to overlap with neighboring sites, breaking the validity of the single-site variational approximation. These results confirm the applicability of the variational model in the tight-binding regime and support the use of $\Delta x(t)$ as a robust observable of internal dynamics.}
	\label{fig:fig5}
\end{figure*}

\section{Variational Model and Analytical Interpretation}
\label{sec:model}

To deepen our understanding of the condensate dynamics during the transport protocol, we develop a simplified analytical model based on two key assumptions: (i) the condensate wave function is well localized within each lattice site, and (ii) the global wave function results from the coherent superposition of identical intra-site wave packets.

This approximation is justified in our regime of deep lattice potentials ($V_0 \simeq 104\,E_r$), where tunneling between sites is negligible and tight-binding conditions are satisfied. As shown in Figure~\ref{fig:fig1}, the spatial extent of the condensate is much larger than the lattice spacing, resulting in the population of more than 800 individual sites. Moreover, the atomic density varies very little over a few lattice periods, making it reasonable to assume, at first approximation, that the intra-site dynamics are identical across the lattice.

We thus write in the moving frame the global wave function as a sum over localized contributions centered at each site
\begin{equation}
\Psi(x, t) \propto \sum_{n} \Phi(x - n\,Q, t),
\end{equation}
where $Q = \lambda_L/2$ is the lattice period, and $\Phi(x, t)$ is the intra-site wave function. This decomposition is justified by the negligible overlap between wave packets localized at different sites. Taking the Fourier transform of this sum yields
\begin{equation}
\bar{\Psi}(k, t) \propto \bar{\Phi}(k, t) \sum_{n} e^{-i n k Q},
\end{equation}
leading at final time $t=t_f$ to the well-known diffraction formula\cite{Garcion_2024,Lecoffre_2025}
\begin{equation}
|\bar{\Psi}(k, t_f)|^2 \propto \left(\frac{\sin(N_{\text{s}}u)}{\sin(u)}\right)^{\!\!2} |\bar{\Phi}(k, t_f)|^2,
\end{equation}
with $N_{\text{s}}$ the number of populated sites and $u = \pi k /(2 k_L)$. This expression reveals a momentum-space structure composed of narrow peaks spaced by $2k_L$, modulated by the global envelope $|\bar{\Phi}(k, t_f)|^2$ computed at the end of the final release, i.e. at $t_f=2t_L+t_\mathrm{acc}$.

To estimate this envelope, we model the intra-site dynamics with a time-dependent variational ansatz for the localized wave function $\Phi(x, t)$ using the Gaussian form
\begin{equation}
\Phi(x, t) = \frac{1}{[\pi\sigma^2(t)]^{\frac{1}{4}}} \exp\!\left(-\frac{x^2}{2\sigma^2(t)}\right),
\label{eq:var-an}
\end{equation}
where $\sigma(t)$ describes the size dynamics of the localized wave packet. Its evolution obeys a non-linear Ermakov-type equation derived from the time-dependent variational principle \cite{Zoller1996, Zoller1997}
\begin{equation}
\ddot{\sigma}(t) + \omega_{\mathrm{OL}}^2(t)\,\sigma(t) =
\frac{\hbar^2}{4 m^2 \sigma^3(t)} +
\frac{\hbar\omega_{\perp} a_s N_{}}{4\sqrt{\pi}\,m\,\sigma^2(t)},
\label{eq:var}
\end{equation}
with $\omega_{\mathrm{OL}}(t)$ the time-dependent trap frequency within each lattice site. In the non-interacting case, the last term vanishes, and $\sigma(t)$ evolves then solely under the balance between confinement, \emph{i.e.} the restoring force from the trapping potential represented by the term $\omega_{\text{OL}}^2(t)\sigma(t)$, and quantum pressure, which arises from Heisenberg's uncertainty principle and drives wave packet expansion through the term $\hbar^2 / (4m^2\sigma^3(t))$ \cite{Pethick2008}.

\begin{figure*}[t!]
	\centering
	\includegraphics*[width=1.75\columnwidth]{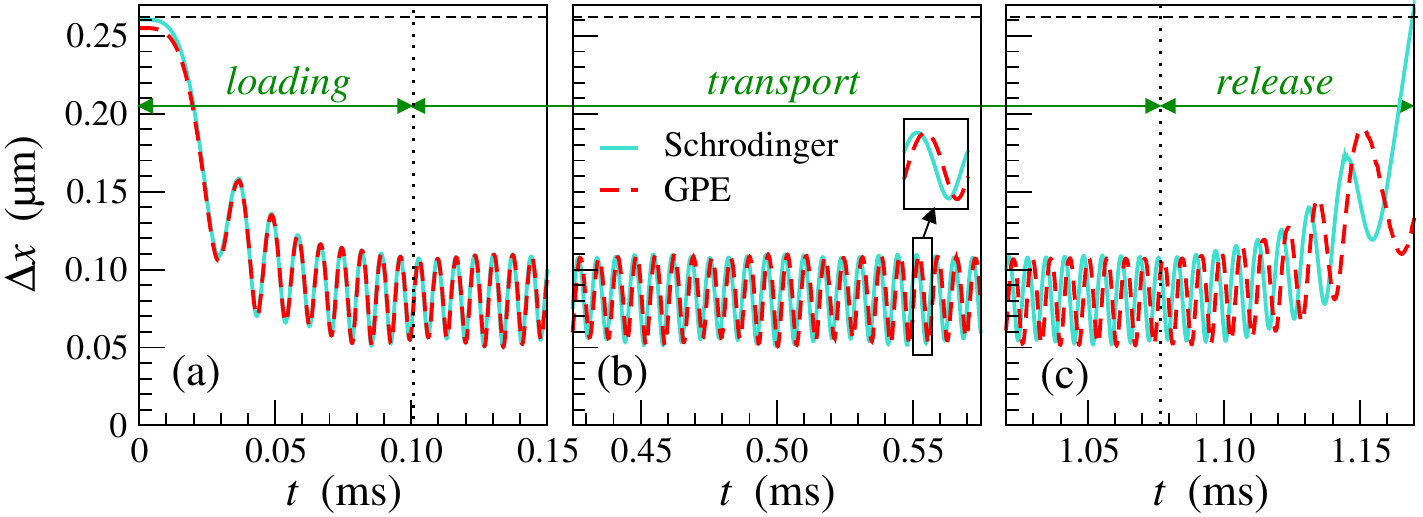}
	\caption{Impact of interactions on the intra-site dynamics of the condensate wave packet. The time evolution of the spatial width $\Delta x(t)$ (FWHM) is shown for $t_L = 100~\mu\mathrm{s}$ and $t_{\mathrm{acc}} = 976.6~\mu\mathrm{s}$, using two models: the full Gross–Pitaevskii equation (dashed red line) and the linear Schr\"odinger equation (solid turquoise line). Both simulations start with a similar width, $\Delta x(0) \simeq 0.26~\mu\mathrm{m}$. During the loading phase (a), the dynamics remain indistinguishable, indicating that interactions play a negligible role at early times. However, a slight dephasing emerges near the end of the acceleration phase (b), as revealed in the central inset, and becomes more pronounced during the release stage (c). This divergence stems from the repulsive interactions included in the GPE, which tend to alter the wave packet breathing dynamics. Notably, in this example, the interacting solution yields a narrower final width, corresponding to a broader momentum distribution and degraded spectral purity. These results illustrate how interactions can act as a source of dephasing and, depending on the dynamical context, either improve or degrade the monochromaticity of the transported condensate under non-adiabatic conditions.}
	\label{fig:fig6}
\end{figure*}

Once $\sigma(t)$ is computed by numerically solving Eq.\,(\ref{eq:var}) using a fifth-order Runge–Kutta method, the momentum-space envelope can be evaluated as
\begin{equation}
|\bar{\Phi}(k, t_f)|^2 \propto \exp[-\sigma^2(t_f) k^2],
\end{equation}
directly linking the final intra-site width $\sigma(t_f)$ to the spectral content of the wave function. This model provides the theoretical foundation for understanding the magic times observed in Section \ref{sec:results}, where optimal spectral purity coincides with specific phases of the breathing oscillation.

\section{Monochromaticity}
\label{sec:monochromaticity}

Building on the magic times identified in Section \ref{sec:results}, we now examine the physical mechanism underlying this phenomenon through the lens of our variational model. Figure~\ref{fig:fig5} compares the evolution of the intra-site density FWHM, $\Delta x(t)$, computed via this variational model (solid black line), with full numerical results obtained from the Gross–Pitaevskii equation (dashed red lines), for an acceleration time of $t_\mathrm{acc} = 976.6~\mu\mathrm{s}$. The variational propagation is initialized at the time $t_i$ of the fourth local maximum of $\Delta x(t)$, as obtained from the GPE solution. The corresponding value of $\sigma(t_i)$ is set from the numerical result to ensure accurate matching at the starting point. The agreement is excellent during the second half of the loading stage and throughout the transport. This confirms the validity of the variational ansatz (\ref{eq:var-an}) in the tight-binding regime, where the condensate remains well localized within individual lattice sites. Discrepancies emerge only after release, when the wave function spreads and develops non-Gaussian or inter-site overlapping features in free space. This reflects the breakdown of the local harmonic approximation underlying the variational model, and the inability of the Gaussian ansatz (\ref{eq:var-an}) to capture the resulting overlap dynamics.

These results confirm however that the oscillations observed in the momentum populations $P_0$, $P_{-2}$ and $P_{+2}$ are driven by internal breathing dynamics, rather than inter-site interference. The model also provides an intuitive explanation for the link between the final value of $\Delta x(t)$ and the spectral purity of the wave packet since broader final spatial profiles correspond to narrower momentum distributions, and vice versa. $\Delta x(t_f)$ is thus linked by Fourier transform to the envelope of the momentum distribution $P(k)$. As a result, changes in $\Delta x(t_f)$ directly translate into modulations of the population in the central and sideband momentum peaks, $P_0$ and $P_{\pm2}$, as illustrated in Figure~\ref{fig:fig3}.

These breathing-like size oscillations provide a clear physical origin for the periodic population transfer observed in momentum space since this coherent modulation of the wave packet’s spatial extent naturally leads to population redistribution among momentum modes. As such, the mechanism, now fully characterized, offers a powerful strategy for controlling the spectral purity (or monochromaticity) of the transport procedure, even under fast loading protocols, such as the one implemented here with $t_L = 100~\mu\mathrm{s}$.

We have verified that the magic-time phenomenon is robust with respect to the detailed shape of the acceleration ramp. Replacing the linear ramps of the trapezoidal profile by sinus-squared ramps, which ensure continuity of the time derivative of the acceleration, yields identical magic times with identical efficiencies. This observation confirms that the underlying breathing-based mechanism depends primarily on the total acceleration duration rather than on the specific temporal profile, and indicates that the choice of a trapezoidal ramp in this study should not be viewed as a practical limitation.

Having clarified the underlying dynamics using both the variational model and Gross–Pitaevskii simulations, we now turn to examine the role of interactions more closely. To this end, we compare in Fig.~\ref{fig:fig6} the full Gross–Pitaevskii equation (which includes mean-field interactions) with the linear Schrödinger equation (obtained by setting $g_{1D} = 0$), both solved numerically without any variational approximation. This comparison, distinct from the variational analysis of Fig.~\ref{fig:fig5}, isolates the specific contribution of atomic interactions to the breathing dynamics. Figure~\ref{fig:fig6} thus presents a direct comparison of the time-dependent intra-site size dynamics $\Delta x(t)$ obtained from the full Gross–Pitaevskii equation (dashed red line) and from the linear Schr\"odinger equation (solid turquoise line), using identical loading and acceleration parameters: \(t_L = 100~\mu\mathrm{s}\) and \(t_{\mathrm{acc}} = 976.6~\mu\mathrm{s}\).

Initially, the dynamics starts with a similar intra-site spatial width, $\Delta x(0) \simeq 0.26~\mu\mathrm{m}$. The loading and acceleration parameters are fixed at $t_L = 100~\mu\mathrm{s}$ and $t_{\mathrm{acc}} = 976.6~\mu\mathrm{s}$, corresponding to a configuration previously shown to yield a polychromatic final state (see Fig.~\ref{fig:fig3}). During the loading phase, both models produce nearly identical breathing oscillations, indicating that interactions have a negligible effect in this initial stage. A slight dephasing begins to appear near the end of the acceleration stage, and becomes more pronounced during the release phase. The subplot (c) reveals this divergence clearly, with the GPE solution retaining a smaller spatial width at final time compared to the non-interacting case. This final spatial compression observed in this particular case for the interacting system implies a broader momentum distribution, and thus degraded spectral purity. These results demonstrate that, under specific conditions, repulsive interactions can modify non-adiabatic excitations and can degrade or help restore monochromaticity depending on the dynamical context. Accounting for these effects is essential for coherent matter-wave transport and interferometry applications, where final-state purity and stability are paramount.

\begin{figure}[t!]
\centering
\includegraphics*[width=0.99\columnwidth]{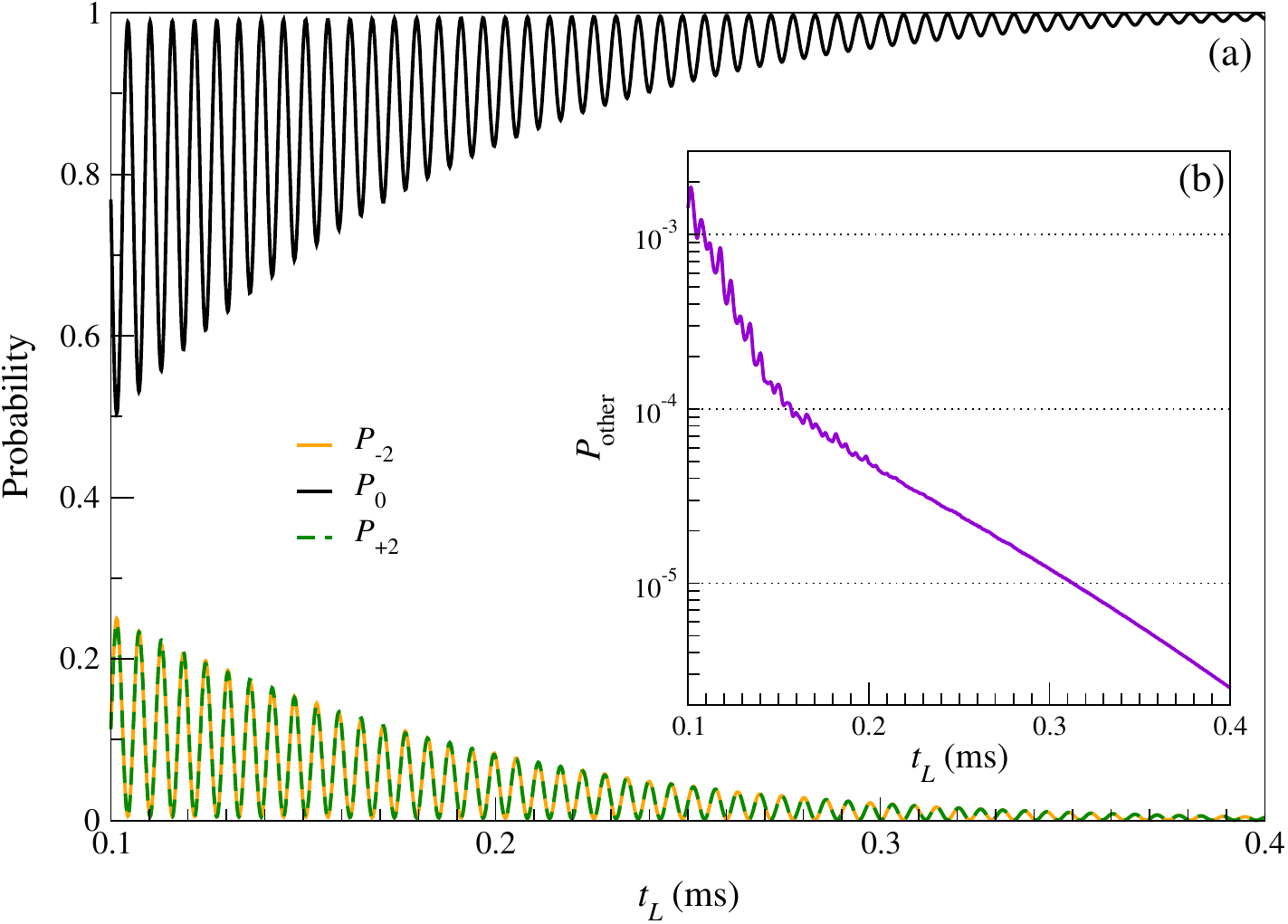}
\caption{Final momentum-state populations as a function of the lattice loading and release time $t_L$, for a fixed acceleration duration $t_{\mathrm{acc}} = 0.7~\mathrm{ms}$. The main panel shows the populations in the principal momentum classes: central peak $P_0$ (black solid line), and first-order sidebands $P_{-2}$ (orange solid line) and $P_{+2}$ (green dashed line). For short loading times ($t_L \lesssim 0.3~\mathrm{ms}$), pronounced oscillations are observed, with a period of approximately $6\,\mu\mathrm{s}$. These oscillations exhibit damped beating patterns, reflecting a decreasing transfer efficiency. Indeed, as $t_L$ increases, the central component $P_0$ displays oscillations converging toward unity, while the sideband populations $P_{\pm 2}$ decay rapidly toward zero, indicating the suppression of higher-band excitations and the recovery of adiabaticity. The inset shows the residual population in all other momentum states, $P_{\mathrm{other}}$, plotted on a logarithmic scale. Its exponential decay with $t_L$ confirms the progressive refinement of the spectral profile. Since the same envelope is used for both loading and release, longer $t_L$ values ensure not only smoother initialization, but also minimize excitations during the final projection into free space.}
\label{fig:fig7}
\end{figure}

Figure~\ref{fig:fig7} finally shows the dependence of the final momentum-state populations on the lattice loading time $t_L$, for a fixed acceleration duration $t_{\mathrm{acc}} = 0.7~\mathrm{ms}$. The main panel displays the populations in the primary diffraction orders: the central momentum component $P_0$ (black solid line), and the first-order sidebands $P_{-2}$ (orange solid line) and $P_{+2}$ (green dashed line).

For short loading durations ($t_L \lesssim 0.3~\mathrm{ms}$), all three curves exhibit pronounced oscillations with a characteristic period of approximately $6\,\mu\mathrm{s}$. These modulations arise from coherent breathing dynamics within the lattice sites, associated with the population of multiple momentum components during the non-adiabatic loading and/or release stages. The resulting beating patterns reflect the imperfect projection of the condensate into a single momentum class.

As $t_L$ increases, the central population $P_0$ rises steadily and progressively converges toward unity. This trend signals the suppression of coherent excitations and a transition toward adiabatic loading. The rapid decay of the amplitude of oscillations with increasing $t_L$ indicates a reduced population transfer among momentum modes and improved spectral purity.

The sideband populations $P_{-2}$ and $P_{+2}$ exhibit complementary damped oscillations, both tending to zero as $t_L$ increases. For instance, at $t_L \approx 0.1~\mathrm{ms}$, $P_{-2}$ reaches values as high as 0.2, but rapidly decreases as the system transitions toward a single-momentum state. Even though the symmetry between $P_{-2}$ and $P_{+2}$ reflects the time-reversal invariance of the symmetric acceleration profile, this condition alone does not guarantee exact equality. Perfect symmetry also requires that the potential within each lattice site remains spatially symmetric throughout the sequence. In the co-moving frame, the inertial term $m\,a_\mathrm{OL}(t)\,\hat{x}$ in Eq.~(\ref{eq:GPE2}) breaks this symmetry by introducing a directional tilt during the acceleration phase. As a result, a small imbalance between $P_{-2}$ and $P_{+2}$ can appear, especially for short acceleration durations, where $a_{\max}$ is large. In the present case, the effect is negligible due to the relatively long plateau durations used, which keep $a_{\max}$ moderate and the induced asymmetry weak.

The inset shows the total population in all higher-order momentum states, $P_\mathrm{other}$, plotted on a logarithmic scale. Its exponential decay confirms the rapid suppression of high excitations and the emergence of a spectrally pure, nearly monochromatic wave packet. For $t_L \gtrsim 0.3~\mathrm{ms}$, $P_\mathrm{other}$ drops below $10^{-5}$ while $P_{-2}$ and $P_{+2}$ remain negligible, reflecting a state of high spectral purity.

To quantify the practical advantages of the magic-time protocol, we now analyze the speedup gain over adiabatic dynamics as a function of lattice depth. The adiabaticity of the transport process is controlled by two main timescales: the loading/release duration $t_L$ and the acceleration duration $t_{\rm acc}$, the total duration of the sequence being $t_{\rm tot} = 2 t_L + t_{\rm acc}$. We define the adiabatic limit as the regime where the population in the target momentum state exceeds 98\%, ensuring that sideband populations remain negligible with $P_{\pm 2} < 1\%$ each.

\begin{table*}[t]
\centering
\caption{Adiabatic $t_{\rm acc}^{\rm ad}$ and magic $t_{\rm acc}^{\rm mag}$ acceleration times for different lattice depths. The total adiabatic $t_{\rm tot}^{\rm ad}$ and magic $t_{\rm tot}^{\rm mag}$ durations of the full loading + acceleration + release sequence are also given, as well as the speedup factors $t_{\rm tot}^{\rm ad}/t_{\rm tot}^{\rm mag}$. For the adiabatic protocol, the loading/release time is fixed at $t_L = 1$\,ms, while it is fixed at $t_L = 100\,\mu$s for the optimized protocol.\\}
\label{tab:speedup}
\setlength{\tabcolsep}{12pt}
\begin{tabular}{c|c||c|c||c|c||c}
& & \multicolumn{2}{c||}{Adiabatic Protocol} & \multicolumn{2}{c||}{Optimized Protocol} & \multirow{2}{*}{Speedup}\\
Laser     & Lattice depth & \multicolumn{2}{c||}{$(t_L = 1\,\mathrm{ms})$} & \multicolumn{2}{c||}{$(t_L = 100\,\mu\mathrm{s})$} & \\[3pt]
Power (W) & $V_0/E_r$ & $t_{\rm acc}^{\rm ad}$ ($\mu$s) & $t_{\rm tot}^{\rm ad}$ ($\mu$s) & $t_{\rm acc}^{\rm mag}$ ($\mu$s) & $t_{\rm tot}^{\rm mag}$ ($\mu$s) & $t_{\rm tot}^{\rm ad}/t_{\rm tot}^{\rm mag}$\\[3pt]
\hline
\rule{0pt}{12pt}\!\!
2 &  52 & 380 & 2\,380 & --- & --- & --- \\
3 &  78 & 275 & 2\,275 & 560 & 760 & $\times$ 3.0 \\
4 & 104 & 230 & 2\,230 & 300 & 500 & $\times$ 4.4 \\
5 & 130 & 200 & 2\,200 & 230 & 430 & $\times$ 5.1 \\
6 & 156 & 180 & 2\,180 & 165 & 365 & $\times$ 6.0 \\
\end{tabular}
\end{table*}

As shown in Table~\ref{tab:speedup}, for purely adiabatic loading/release with $t_L = 1$\,ms, the minimum acceleration duration $t_{\rm acc}^{\rm ad}$ required to reach this threshold decreases with lattice depth, from 380\,$\mu$s at $V_0 = 52\,E_r$ to 180\,$\mu$s at $V_0 = 156\,E_r$. A fully adiabatic protocol thus requires a total duration $t_{\rm tot}^{\rm ad}$ varying between about 2.2 and 2.4\,ms. In contrast, the magic-time protocol enables operation in the strongly non-adiabatic regime. The table lists the first magic acceleration times $t_{\rm acc}^{\rm mag}$ achieving $>98\%$ population transfer for a fixed non-adiabatic loading time $t_L = 100\,\mu$s. Importantly, magic times arise periodically, synchronized with the intra-site breathing oscillation period (in the range 5--10\,$\mu$s), offering experimental flexibility in choosing the optimal acceleration duration.

The speedup gain of the optimized protocol relative to the adiabatic protocol, defined as the ratio of total sequence durations $t_{\rm tot}^{\rm ad}/t_{\rm tot}^{\rm mag}$, is shown in the last column of Table~\ref{tab:speedup}. This gain increases with lattice depth, reaching factors of 3 to 6. However, the magic-time mechanism relies on coherent intra-site breathing dynamics and therefore requires operation in the tight-binding regime where $V_0 \gg E_r$. At shallow lattice depths, tunneling between neighboring sites becomes significant during the acceleration phase, disrupting the periodic breathing oscillations. This is evidenced by the behavior at the shallowest lattice depth considered here ($V_0 \simeq 52\,E_r$), where the 98\% fidelity threshold is never reached using magic times. Based on our simulations, for a loading/release duration $t_L$ of 100\,$\mu$s, the protocol remains effective for lattice depths $V_0 \gtrsim 70\,E_r$, setting a practical lower bound for its applicability. Importantly, this tight-binding regime is precisely where future large-momentum-transfer interferometry is heading, as achieving transfers of hundreds to thousands of $\hbar k_L$ requires deep optical lattices to maintain atomic confinement throughout the acceleration sequence.

Finally, it is important to note that the loading time $t_L$ also defines the timescale of the final release stage, since the same envelope function is used. Hence, increasing $t_L$ not only improves the initialization of the system, but also ensures minimal excitation during the final release, ultimately leading to the wave packet projection into free space, highlighting the dual role of the parameter $t_L$ in this particular transport scheme. Remarkably, and analogous to the magic times identified for $t_{\mathrm{acc}}$ in Section \ref{sec:results}, we find that specific loading durations $t_L$ also exhibit optimal values where the population $P_0$ approaches unity despite the non-adiabatic ramp. These magic loading times, like their acceleration counterparts, arise from synchronization with the breathing period, where destructive interference between transiently populated $\pm 2\,\hbar k_L$ momentum states effectively rephases the wave function into a narrow final momentum distribution. This demonstrates that the magic time phenomenon is a general feature of the transport protocol, applicable to both the loading/release and acceleration stages, enabling spectrally pure wave packets without requiring adiabatic protocols or complex control fields.\\

\section{Conclusion}
\label{sec:conclusion}

In this work, we have presented a comprehensive numerical investigation of the accelerated transport of a Bose–Einstein condensate in a one-dimensional optical lattice. Using the time-dependent Gross–Pitaevskii equation, we modeled the full transport sequence, comprising (i) the preparation of the initial ground state in a harmonic trap, (ii) loading into the optical lattice via fast ramps, (iii) coherent acceleration using a symmetric trapezoidal profile, and (iv) fast release into free space. The protocol was designed to impart a precisely controlled momentum to the condensate while preserving its internal coherence.

Our simulations reveal the critical importance of the loading, acceleration and release timescales in determining the final momentum distribution. In particular, we identified the interplay between intra-site breathing dynamics and momentum redistribution as the dominant mechanism limiting spectral purity. The identification of magic times, \emph{i.e} specific loading/release or acceleration durations synchronized with the breathing period, provides a clear operational principle for achieving coherent momentum-selective transport even in the non-adiabatic regime. The emergence of narrow, spectrally pure momentum distributions was indeed shown to correlate with maxima of breathing oscillations in the final condensate's spatial width. Although our simulations were performed with 190 momentum kicks (a value relevant to compact fountain geometries \cite{Abend_2016}), the periodic nature of the breathing dynamics suggests that the magic-time mechanism should scale to larger momentum transfers without qualitative change, extending the applicability of our protocol to high-baseline interferometry.

We further demonstrated that the final momentum state transfer efficiency can be inferred from real-space observables such as the intra-site width of the condensate, which serves as a sensitive probe of spectral purity under non-adiabatic loading and release conditions. A variational model provided quantitative agreement with the full Gross–Pitaevskii dynamics during the loading and acceleration stages, and we highlighted the role of interactions in modulating the breathing dynamics.

The framework developed here offers both diagnostic and predictive power for optimizing coherent transport protocols in optical lattices. These findings are directly applicable to quantum sensing platforms where timing constraints limit the use of traditional adiabatic protocols. The identification of “magic times”, at which spectral purity is maximized despite rapid loading and release dynamics, demonstrates that the usual trade-off between speed and monochromaticity can be circumvented. This opens a practical route toward the rapid generation of coherent matter-wave sources ideally suited for compact and time-sensitive interferometric architectures without complex control protocols. From a practical standpoint, the relevant timescales remain accessible to standard experimental control systems. The breathing period scales as $T_{\text{br}} \propto 1/\sqrt{V_0}$, and for typical $^{87}$Rb experiments in the tight-binding regime, it lies in the range of 5 to 10 $\mu$s depending on laser intensity. Crucially, sub-microsecond timing precision is not required: since magic times correspond to extrema of the breathing oscillation, the system is inherently robust to small timing variations near these optimal points. Moreover, the condition for high spectral purity is satisfied over a finite time window around each magic time, rather than at a single precise instant, as the relevant criterion is that the final wave packet be sufficiently broad in real space (and hence narrow in momentum space) to suppress sideband populations. Additional flexibility is provided by the loading duration $t_L$, which controls the amplitude of breathing oscillations and can be adjusted to further relax timing constraints when needed.

An important aspect not addressed in this work is the interferometric phase accumulated during the transport sequence. For applications in precision atom interferometry, understanding how this phase depends on ramp profiles, lattice depth, and the choice of magic times would require additional investigations. Recent theoretical frameworks have been developed to accurately describe phase accumulation in Bloch-oscillation-enhanced atom interferometry \cite{Fitzek_2024}, while experimental techniques have been demonstrated to characterize these phases \cite{Rahman_2024}. Furthermore, optimal control techniques have recently shown promising results for Bragg-based interferometry in theoretical studies, including robust high-efficiency beam splitters \cite{RuiLi_2024}, high-contrast interferometers \cite{RuiLi_2025}, and diffraction-phase-free schemes \cite{Lahuerta_2025}, suggesting that similar approaches could be extended to Bloch oscillation sequences. Extending the present analysis to include a systematic study of the phase behavior at magic times, building on these recent advances, represents a natural direction for future work.

Furthermore, while the present study focuses on Bose-Einstein condensates, our results suggest that the magic-time mechanism should extend to thermal atomic clouds. Indeed, we have demonstrated that the protocol works in two limiting cases: interacting condensates (via the Gross-Pitaevskii equation) and non-interacting atoms (via the Schrödinger equation). Breathing oscillations in thermal clouds have been experimentally observed \cite{Lobser_2015} and can be described by generalized scaling equations that continuously interpolate between these two validated limits \cite{DGO_2002, Pedri_2003}. Since the magic-time phenomenon relies on synchronization with breathing dynamics, and since breathing modes persist across this entire range of interaction strengths, we expect the mechanism to remain operative in the intermediate thermal regime, with oscillation periods that depend on the ratio of thermal to mean-field energy \cite{DGO_2002, Pedri_2003}.

Beyond its practical implications for atom interferometry, state preparation, and quantum metrology, this approach provides a foundation for exploring more complex regimes involving disorder, strong interactions, or higher dimensionality. The interplay between transport, non-adiabatic excitation and momentum transfer in these systems presents rich physics that can be investigated with the framework established here. Our work thus provides both an immediately implementable protocol for current quantum sensing applications and a robust platform for fundamental studies of non-equilibrium quantum dynamics.

\section*{Dedication}
We are pleased to contribute this article to the special issue of \emph{AVS Quantum Science} honoring the 60$^{th}$ birthday of Ernst Maria Rasel, whose pioneering work has profoundly shaped the field of cold atom physics. Happy Birthday Ernst!

\section*{Acknowledgments}
N.G. acknowledges funding from the Deutsche Forschungsgemeinschaft (German Research Foundation) under Germany’s Excellence Strategy (EXC-2123 QuantumFrontiers Grants No. 390837967) and Project-ID 274200144– SFB 1227 (DQ-mat, project A05), as well as funding by the AGAPES project - grant No 530096754 within the ANR-DFG 2023 Programme.

\section*{Author Declarations}
\subsection*{Conflict of Interest}
The authors have no conflicts to disclose.

\section*{Data Availability}
The data that support the findings of this study are available from the corresponding author upon reasonable request.\\

\bibliography{bibliography}

\end{document}